\documentclass[12pt]{article}

\global\arraycolsep=1pt

\usepackage{color,amsbsy,amssymb,latexsym,amsfonts, amsmath,cancel}
\usepackage{braket}
\usepackage{mathrsfs}
\usepackage{graphicx}

\newcommand{\bel}[1]{\begin{equation}\label{#1}}                     
\newcommand{\bal}[1]{\begin{eqnarray}\label{#1}}   
\newcommand{\be}{\begin{equation}}               
\newcommand{\ba}{\begin{eqnarray}}           
\newcommand{\ee}{\end{equation}}
\newcommand{\ea}{\end{eqnarray}}

\renewcommand{\thefootnote}{\fnsymbol{footnote}}

\newcommand{\bea}{\begin{equation}}
\newcommand{\eea}{\end{equation}}

\begin{document}

%
%
\begin{titlepage}
\begin{flushright}
\normalsize
~~~~
OCU-PHYS 356\\
July, 2012
\end{flushright}

\vspace{15pt}

\begin{center}
{\LARGE $D$-term Dynamical Supersymmetry Breaking }\\
\vspace{5pt}
{\LARGE  Generating Split ${\cal N} =2$ Gaugino Masses }\\
\vspace{5pt}
{\LARGE of Mixed Majorana-Dirac Type } \\
\end{center}

\vspace{23pt}

\begin{center}
{ H. Itoyama$^{a, b} $\footnote{e-mail: itoyama@sci.osaka-cu.ac.jp}
  and 
 Nobuhito Maru$^c$\footnote{e-mail: maru@phys-h.keio.ac.jp} 
}\\
%
\vspace{18pt}
%

$^a$ \it Department of Mathematics and Physics, Graduate School of Science\\
Osaka City University  and\\
\vspace{5pt}

$^b$ \it Osaka City University Advanced Mathematical Institute (OCAMI) \\

\vspace{5pt}

3-3-138, Sugimoto, Sumiyoshi-ku, Osaka, 558-8585, Japan \\

$^c$ \it Department of Physics, and
Research and Education Center for Natural Sciences, \\
Keio University, Hiyoshi, Yokohama, 223-8521 JAPAN

\end{center}
%
\vspace{20pt}
\begin{center}
Abstract\\
\end{center}
%
Under a few mild assumptions,  
${\cal N}=1$  supersymmetry in four dimensions is shown to be spontaneously broken in a metastable vacuum 
 in a self-consistent Hartree-Fock approximation of BCS/NJL type to the leading order, 
 in the gauge theory specified by the gauge kinetic function and the superpotential of adjoint chiral superfields, 
 in particular,  that possesses ${\cal N}=2$ extended supersymmetry spontaneously broken to ${\cal N} =1$ at tree level.
We derive an explicit form of the gap equation, showing the existence of a nontrivial solution.
The ${\cal N}=2$ gauginos in the observable sector receive mixed Majorana-Dirac masses and are split
 due to both the non-vanishing $\langle D^0 \rangle$ and $\langle F^0 \rangle$ induced with $\langle D^0 \rangle$.
  It is argued that proper physical applications and assessment of the range of the validity of our framenwork
  are made
  possible by rendering the approximation into  $\frac{1}{N^2}$ expansion.


\vfill

\end{titlepage}

\renewcommand{\thefootnote}{\arabic{footnote}}
\setcounter{footnote}{0}

\section{introduction}
Identifying  the mechanism of spontaneously broken ${\cal N}= 1$ supersymmetry 
 in nature has been the vital issue in theoretical particle physics for three decades.
 There are two order parameters in supersymmetric field theories 
 which tell whether ${\cal N} =1$ supersymmetry is spontaneously broken or not. 
 The one is  the nonvanishing vacuum expectation value (vev for short) of
  the $F$ term, the auxiliary field of the chiral superfield. 
There are mechanisms known that generate the nonvanishing $F$ term 
 both at tree (classical)\cite{Oraif} and quantum mechanical levels. 
As the nonrenormalization theorem applies \cite{GRS} here, 
 the consideration beyond the lowest order in perturbation theory 
 ought to be genuinely nonperturbative in nature.
 Instantons have played important roles in the development of this investigation 
 in the past years \cite{Witten, ADS, Veneetal}.
  The other is the nonvanishing vev of the $D$ term, the auxiliary field of the vector superfield.
  There is a well-known mechanism at the tree level \cite{FI} to this and the nonrenormalization theorem
   does not apply.   In the past, there were models found, (for example, the $4-1$ model)\cite{DNNSCFK}
    which display dynamical supersymmetry breaking triggered by the nonperturbative effects lying 
    in the suprepotential. They generate both F-terms and D-terms.  
  
In this letter, we will provide 
 mechanism that provides dynamical supersymmetry breaking triggered by a
 nonvanishing  $D$-term 
 beyond the lowest order in perturbation theory.
 We will treat ${\cal N}=1$ effective theory characterized by three nontirivial input functions, a Kahler potential,
  a gauge kinetic funtion and  a superpotential.  The vacuum of the theory at the tree level
   is assumed to preserve ${\cal N}=1$ supersymmetry; neither $D$ term nor $F$ term is generated at the
   tree level.
 Our mechanism assumes the existence of scalar gluons in nature and, in that respect, is distinct from
  the previous proposals on dynamical supersymmetry breaking.  Our mechanis m 
 involves condensation of composite operators
  and is based on the self-consistent Hartree-Fock approximation
 which is reminiscent of that of \cite{BCS,NambuPRBCS,NJL}
 in the theory of superconductivity/chiral symmetry.
 (For the ideas in the past of using a NJL type of approach to assess dynamical
supersymmetry breaking, gaugino condensation  and  the comparison with computation from instantons
 and symmetries, see, for instance, \cite{MRFJKBL}.)
  Such a possibility is precluded by the requirement of the perturbative renormalizability of 
  the interactions in the lagrangian. 
 With the advent of UV finite string models and several experimental calls beyond the standard model
 in particle physics, however, the mechanism in what follows can be made relevant   
 in the effective field theory description for
 the energy scale up to the multi TEV that is being probed by the Large Hadron Collider. 
 The original $D=0$ perturbative vacuum is not lifted in our treatment 
and the new local minimum we find is metastable.  As is anticipated from the study
 of the models \cite{DNNSCFK} mentioned, our mechanism eventually generates  a nonvanishing F term as well.

\section{basic observation and mechanism}
 Let us start from a general lagrangian 
     \ba
    {\cal L}
     &=&     
           \int d^4 \theta K(\Phi^a, \bar{\Phi}^a)
+ (gauging)  
+ \int d^2 \theta
           {\rm Im} \frac{1}{2} 
           \tau_{ab}(\Phi^a)
           {\cal W}^{\alpha a} {\cal W}^b_{\alpha}
+ \left(\int d^2 \theta W(\Phi^a)
         + c.c. \right),      
           \label{KtauW}
    \ea
 where $K$ is a K\"ahler potential with its gauging by the gauge group understood, 
 $\tau_{ab}(\Phi^a)$ is a gauge kinetic superfield of the chiral superfield $\Phi^a$ 
 in the adjoint representation, and $W(\Phi^a)$ is a superpotential. 

 The bilinears made of the ${\cal N} = 1$ gaugino $\lambda^a$ and the matter fermion $\psi^a$, 
 which are referred to as ${\cal N}=2$ gauginos in this letter, 
 are obtained from the second and the third line of eq.(\ref{KtauW}): 
\ba
-\frac{1}{2} (\lambda^a, \psi^a) 
\left(
\begin{array}{cc}
0 & -\frac{\sqrt{2}}{4} \tau_{abc} D^b \\
-\frac{\sqrt{2}}{4} \tau_{abc} D^b & \partial_a \partial_c W \\
\end{array}
\right) 
\left(
\begin{array}{c}
\lambda^c \\
\psi^c \\
\end{array}
\right) + (c.c.), 
\label{Majorana-Dirac}
\ea 
 where $\tau_{abc} \equiv \partial_c \tau_{ab}$.
 Note that the nonvanishing value of $\tau_{abc}$ ensures the coupling of the auxiliary $D^a$ field
   to the fermionic bilinears while there is no bosonic  counterpart in the lagrangian.
 Let us assume that $\tau_{ab}$ is obtained as the second derivatives of a trace function $f(\Phi^a)$.
  The nonvanishing vevs are $\langle \tau_{0aa} \rangle$.
 The holomorphic and nonvanishing part of the mass matrix is
\ba
  M_{F, a} \equiv
\left(
\begin{array}{cc}
0 & -\frac{\sqrt{2}}{4} \langle \tau_{0aa} D^0 \rangle \\
-\frac{\sqrt{2}}{4} \langle \tau_{0aa} D^0 \rangle & \langle \partial_a \partial_a W \rangle \\
\end{array}
\right) 
\label{Fmassmat}
\ea 
 to each generator. We see that, upon diagonalization, this has two unequal and nonvanishing eigenvalues
  provided  $ \langle D^0 \rangle \neq 0$ and
 $\langle \partial_a \partial_a W \rangle \neq 0$.
  In this case, the ${\cal N}=2$ gauginos receive masses of mixed Majorana-Dirac type and are split.
 This observation generalizes the proposal of \cite{FNW} where the masses are
  of a pure Dirac type and the ${\cal N}=2$ gauginos are degenerate while the supersymmetry is broken.
   
  The value of $\langle D^0 \rangle$ is well-determined quantum mechanically 
 in a self-consistent Hartree-Fock approximation as long as the fluctuations are small. 
To one-loop order off-shell, it is given by the stationary condition 
     to the part of the effective potential which contains $D^a$
   \ba
   V^{(D)} + V_{{\rm c.t.}} + V_{{\rm 1-loop}},
   \label{effpot}
   \ea
  where $V^{(D)} = -  \frac{1}{2}g_{ab} D^a D^b$, $g_{ab} = {\rm Im} \tau_{ab}$, $V_{{\rm 1-loop}}$
   is the one-loop contribution and  $V_{{\rm c.t.}}$ is a counterterm 
   which we prepare with regard to the lagrangian.
Note that equation of motion for $D^0$
  tells us that the condensation of the Dirac bilinears is responsible 
  for the nonvanishing order parameter:
$ \langle D^{0} \rangle
    =     - \frac{1}{2 \sqrt{2}} \langle g^{00} 
           \left( \tau_{0cd}\psi^d \lambda^c 
         + \bar{\tau}_{0cd} \bar{\psi}^d \bar{\lambda}^c \right) \rangle$.
   In fact, the stationary condition of eq. (\ref{effpot}) 
  with respect to the auxiliary field $\langle D^0 \rangle $ 
  is nothing but the well-known gap equation of the theory on-shell 
  which contains four-fermi interactions. 

 In order to make our dynamical framework more concrete 
  and to reduce the number of input functions to one,
 we temporarily impose ${\cal N}=2$ supersymmetry on the abelian \cite{APT}
  and nonabelian \cite{FIS12} action. 
For definiteness, we take the gauge group to be $U(N)$ which are unbroken 
 although our results are applicable to a variety of product gauge groups 
 which contain an overall $U(1)$  and which are broken in the vacua specified 
 by the tree level potential.

      The theory we work with is given by the lagrangian 
     \ba
    {\cal L}_{U(N)}
     &=&     {\rm Im} 
           \left[ 
           \int d^4 \theta 
           {\rm Tr} 
           \bar{\Phi} e^{ad V} 
           \frac{\partial {\cal F}(\Phi)}{\partial \Phi} 
        \right. 
      \left.  
          + \int d^2 \theta
           \frac{1}{2} 
           \frac{\partial^2 {\cal F}(\Phi)}{\partial \Phi^a \partial \Phi^b}
           {\cal W}^{\alpha a} {\cal W}^b_{\alpha}
           \right] 
         + \left(\int d^2 \theta W(\Phi)
         + c.c. \right),      
           \label{FIS'}
    \ea
 where the superpotential is
    \bea
    W(\Phi)
     =     {\rm Tr} \left( 2 e \Phi
         + m \frac{\partial {\cal F}(\Phi)}{\partial \Phi} 
           \right).
           \label{prepot}
    \eea
    The superpotential  consists of the electric and magnetic Fayet-Iliopoulos terms
    whose representation is obtained by pointing these two three dimensional vectors 
     in a particular direction under the rigid $SU(2)_R$ rotation \cite{IMMFIS5}.
   The bare action reads
 \ba
 S_{{\rm bare}} = S[ {\cal F}] + S_{ {\rm c.t.}},
 \ea
and
 \ba
S[ {\cal F}] = \int d^4x {\cal L}_{U(N)},  \;\;\;\;\;\;\;  
S_{ {\rm c.t.}} = S[ {\cal F}= (\Lambda/2)(\Phi^0)^2   ].
 \ea
 Here, we have introduced  the supersymmetric counterterm $S_{ {\rm c.t.}} = S[ {\cal F}= (\Lambda/2)(\Phi^0)^2]$
 which conforms to the form of the action.  The counterterm plays a specific role in our treatment of 
 isolating UV infinities at the one-loop effective potential
  and the gap equation in later section. More explicitly, the counterterm to the one-loop effective potential
   is written as
 \ba
V_{{\rm c.t.}} = -{\rm Im}   \frac{\Lambda}{2} \int d^2 \theta {\cal W}^{\alpha a}{\cal W}^a_\alpha 
= -{\rm Im}  \frac{\Lambda}{2} (D^0)^2,
\ea
 where $\Lambda$ is a complex parameter that we have prepared in order to cancel the infinity we will come across.
 
 The effective potential up to one-loop order is
 \ba 
V = V^{(D)} + V^{({\rm sup})} + V_{{\rm 1-loop}} + V_{\rm c.t.},  
\ea
 where $
V^{({\rm sup})} = g^{ab} \partial_a W \overline{\partial_b W}$
and  $\partial_a W = \sqrt{2N} (e \delta_a^0 + m {\cal F}_{0a})$.
 At the tree level, $\langle D^a \rangle =0$ and 
 the vacuum condition is $ \partial_a V^{({\rm sup})} =0 $.
There are two vacua and the positive definiteness of the K\"ahler metric selects one.
We will assume $\frac{{\rm Im}~e}{m} < 0$ for definiteness, 
 so that the condition is $e \delta_c^0 + m {\cal F}_{0c} = 0.$
The second supersymmetry is spontaneously broken at the tree level \cite{APT,FIS12}.
For the spectrum of this theory at the tree level in various phases, see \cite{FIS346}.

\section{effective action up to one-loop and subtraction of UV infinity}

Coming back to the mass matrix eq. (\ref{Fmassmat}) and the issue, 
 we now compute the two eigenvalues ${\bf \Lambda}_a^{(\pm)}$ for each $a$ to be
 \ba
 \label{eigenvalue}
 {\bf \Lambda}_a^{(\pm)} =  m_{a} \lambda^{(\pm)}, \quad 
\lambda^{(\pm)} \equiv \frac{1}{2}\left( 1 \pm \sqrt{1 +\Delta^2} \right), \quad
  \Delta^2 \equiv \frac{(D^0)^2}{4Nm^2},
\ea
where $m_{a} \equiv  \sqrt{2N} m  \langle g^{aa} {\cal F}_{0aa} \rangle$,
  $m$ is the dimension-two parameter of the magnetic Fayet-Iliopoulos term and 
  $| m_{a} |^2$ are the masses of the scalar gluons which are receiving
   experimental attentions.

 It is now clear that the entire contribution to the 1PI vertex function $i \Gamma_{\rm{1-loop}}$
 is
\ba
&& \int d^4x \sum_a | m_{a} |^4 
\int \frac{d^4 \ell^{\mu}}{(2\pi)^4} \ln
 \left[ \frac{(\lambda^{(+)2} - \ell^2 - i \epsilon) (\lambda^{(-)2} - \ell^2 - i \epsilon)}
 {(1 - \ell^2 - i \epsilon) ( - \ell^2 - i \epsilon)}
  \right].
 \ea 
There is a variety of methods available which regulate and evaluate this expression. 
Here we adopt one which first handles the determinant by the integral
$\log a/b = - \int_0^{\infty} dt (e^{-at} - e^{-bt})/t$ and subsequently
continues the power of $t$ to the number designating the spacetime dimension 
\footnote{This is equivalent to the dimensional
reduction as opposed to the dimensional regularization
 scheme in the Feynman diagram language. In the latter, the spinor traces
 are also evaluated in $d$ dimensions 
  while in the former, they are in four dimensions and supersymmetry is preserved
  at least at one-loop \cite{siegel}.}.
  Let us proceed this way.
 The one-loop contribution to the effective potential is expressed as
\ba
 \frac{1}{\sum_a | m_{a} |^4} V_{{\rm 1-loop}} = 
 \frac{1}{16 \pi^2} \int_{0}^{\infty}
 \frac{dt}{t^3} \left( e^{- \lambda^{(+)2} t} + e^{- \lambda^{(-)2} t} - e^{-t} -1 \right).
\ea
 Continuing $3$ to $1+d/2$ in the case of $d$ dimensional integral for regularization, 
 we obtain 
 \ba
  \frac{1}{32\pi^2} \left[ A(d) \left( \Delta^2 + \frac{1}{8} \Delta^4 \right) 
 - \lambda^{(+)4} \log \lambda^{(+)2}   - \lambda^{(-)4}  \log \lambda^{(-)2}  \right],
  \ea
  where 
 \ba
 A(d) = 3/4 - \gamma + \frac{1}{2-d/2}.
\ea
 
 Now the part of the one-loop effective potential which contains $\Delta$ reads
 \ba
  V_{{\rm 1-loop}}^{(D)} &=& V^{(D)} + V_{{\rm c.t.}} + V_{{\rm 1-loop}}  \nonumber \\ 
   &=& \sum_a | m_{a} |^4 \left[ - \beta \Delta^2 - \Lambda_{{\rm res}} \Delta^2 \right. \nonumber \\
   &&+ \left. \frac{1}{32 \pi^2} \left\{ A(d) \left(\Delta^2 + \frac{1}{8} \Delta^4 \right) 
-   \lambda^{(+)4} \log \lambda^{(+)2}   - \lambda^{(-)4}  \log \lambda^{(-)2} \right\}
  \right],
\label{oneloopexp1} 
 \ea
 where $\beta = \frac{ \langle g_{00} \rangle 2Nm^2}{\sum_a | m_{a} |^4}, \;
  \Lambda_{{\rm res}} = \frac{ ({\rm Im} \Lambda) 2Nm^2}{\sum_a | m_{a} |^4}$.
  
  The procedure preserves supersymmetry as both the regularization and the counterterm $V_{{\rm c.t.}}$
  are supersymmetric.
   In  eq.(\ref{oneloopexp1}), however, the  parameter 
   ${\rm Im} \Lambda$ (or $\Lambda_{{\rm res}}$) is still unrelated
   to our one-loop computation and hence the resulting infinity lying in $A(d)$. We can relate these and absorb 
   the infinity into ${\rm Im} \Lambda$ (or $\Lambda_{{\rm res}}$)
    by imposing one (renormalization) condition.  We adopt the following one:
 \ba
  \frac{1}{\sum_a | m_{a} |^4} \left. \frac{\partial^2 V_{{\rm 1-loop}}^{(D)}}{(\partial \Delta)^2} 
\right|_{\Delta=0} = 2c,
 \ea
  where $c$ is a fixed non-universal number.  $\Lambda_{{\rm res}}$ is now expressible in terms of
    $A(d), c, \beta$ as
  \ba
  \Lambda_{{\rm res}}  = \Lambda_{{\rm res}}(d) =  -\beta -\left(c + \frac{1}{64\pi^2}\right)
    +\frac{1}{32\pi^2}A(d).
  \ea
    Our final expression for $V_{{\rm 1-loop}}^{(D)}$ is
 \ba
 &&\frac{1}{\sum_a | m_{a} |^4}  V^{(D)}_{{\rm 1-loop}}  
   =  \left( c + \frac{1}{64 \pi^2} \right)  \Delta^2  + \Lambda_{{\rm res}}^{\prime}(d) 
   \frac{\Delta^4}{8}
   -\frac{1}{32 \pi^2} 
  \left(  \lambda^{(+)4} \log \lambda^{(+)2}   + \lambda^{(-)4}  \log \lambda^{(-)2} \right), 
  \label{VD1loop}
  \ea
  where $\Lambda_{{\rm res}}^{\prime} (d) \equiv c + \beta + \Lambda_{{\rm res}}(d) +\frac{1}{64\pi^2}$.
  By a change of parametrization of the potential from $\Lambda_{{\rm res}}$ to $c$,
   we have been able to isolate the original infinity into the coefficient of $\Delta^4$, which
   is regarded as the oveall scale of the effective potential and the gap equation.
    A similar treatment is seen, for instance, in \cite{colemanweinberghiGN}
   Our treatment differs from  the original treatment of the NJL model \cite{NJL} on spontaneous chiral symmetry breaking
     which  proceeds solely on the bare theory without counterterms and introduces  the relativistic cuttoff.
     We consider this difference as the difference of physics aiming at: in our case,  the UV physics
     which may underlie our model is set by the prepotential function.

  Our computation has been so far with regard to the ${\cal N}=2$ action (eq. (\ref{FIS'}))
   but it is easy to see that the computation and the final conclusion eq. (\ref{VD1loop})
 are valid with regard to the more general ${\cal N}= 1$ action (eq. (\ref{KtauW})) as well.
 The difference between these two cases is absorbed in the redefinition of $\Delta^2$:
 $ \Delta^2 \equiv \frac{(D^0)^2}{4Nm^2}  \Rightarrow  \Delta^2 \equiv \frac{(D^0)^2}{a}$,
   with $a$ a positive number.
 The formulae in this general setting  are
\ba
{\bf \Lambda}_a^{\pm} = \frac{1}{2} m_a 
\left[
1 \pm \sqrt{1+ \Delta^2} \right], \quad m_a \equiv \langle g^{aa} \partial_a \partial_a W \rangle, \quad 
\Delta^2 \equiv \frac{\langle \tau_{0aa}D^0 \rangle^2}{2 \langle \partial_a \partial_a W \rangle^2}. 
\ea

\section{gap equation and nontrivial solution}
  The gap equation is nothing but the stationary condition of eq. (\ref{VD1loop}) with respect to $\Delta$:
\ba
0 = \frac{\partial V^{(D)}_{{\rm 1-loop}}}{\partial \Delta} 
&=& \Delta \left[ 2 \left( c+\frac{1}{64\pi^2} \right) + \frac{\Lambda'_{{\rm res}}}{2} \Delta^2 
- \frac{1}{32\pi^2} \frac{1}{\sqrt{1+\Delta^2}} 
\left\{ (\lambda^{(+)})^3 (2 \log(\lambda^{(+)})^2 + 1) \right. \right. \nonumber \\
&& \left. \left. - (\lambda^{(-)})^3 (2 \log(\lambda^{(-)})^2 + 1) \right\} 
\right]. 
\ea
The gap equation  has a trivial solution $\Delta=0$ which corresponds to the vacuum of
 unbroken supersymmetry. 
Now our interest is whether the nontrivial solution $\Delta \ne 0$ exists or not. 
The gap equation is transcendental, and we first solve it approximately. 
To begin with, consider the case where the D-term VEV is very small, $\Delta^2 \ll 1$. 
Noting that $\lambda^{(+)} \simeq 1, \lambda^{(-)} \simeq 0$ in this case, 
the gap equation can be approximated as 
\ba
2 \left( c+\frac{1}{64\pi^2} \right) + \frac{\Lambda'_{{\rm res}}}{2} \Delta^2 \simeq 
\frac{1}{32\pi^2} \left(1 + \frac{5}{4} \Delta^2 \right). 
\label{apgap1}
\ea
If $c > 0$, then we have no solution of (\ref{apgap1}) because of $\Lambda'_{{\rm res}}$. 
If $c < 0$, then we have a solution $\Delta^2 \simeq -4c/(\Lambda'_{{\rm res}} - 
\frac{5}{2} \cdot \frac{1}{32\pi^2})$. 

Next, consider the case where the D-term VEV is very large, $\Delta^2 \gg 1$. 
Noting that $\lambda^{(\pm)} \simeq \pm \Delta/2$ in this case, 
the gap equation can be approximated as 
\ba
c+\frac{1}{64\pi^2} + \Lambda'_{{\rm res}} \left( \frac{\Delta}{2} \right)^2 
\simeq \frac{1}{32\pi^2} \left( \frac{\Delta}{2} \right)^2 \log \left( \frac{\Delta}{2} \right)^2, 
\label{apgap2}
\ea
which has a unique nontrivial solution $\Delta \ne 0$ if $\Lambda'_{{\rm res}} > 0$.

We have also checked the existence of the nontrivial solution to the gap equation numerically.  
\begin{figure}[htbp]
 \begin{center}
  \includegraphics[width=70mm]{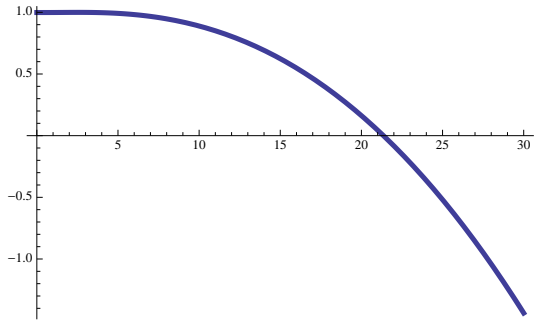}
 \end{center}
 \caption{The plot of the quantity $\partial V_{{\rm 1-loop}}^{(D)}/(\Delta \partial \Delta)$ 
 as a function of $\Delta$. 
 A particular set of parameters $c+\frac{1}{64\pi^2}=1, \Lambda_{{\rm res}}'/8=0.001$ is chosen as an illustration.}
 \label{fig:one}
\end{figure}
In Figure \ref{fig:one}, the quantity $\partial V_{{\rm 1-loop}}^{(D)}/(\Delta \partial \Delta)$ is 
plotted as a function of $\Delta$. 
A unique zero point is found, which implies the existence of a nontrivial solution to the gap equation.  
Although a particular set of parameters is chosen in the Figure \ref{fig:one} to illustrte this, 
we have checked  the existence of a nontrivial solution  in a wide range of parameters.  
Therefore, we conclude that SUSY is broken by the dynamically generated VEV of D-term 
in a self-consistent Hartree-Fock approximation. 

Note that the value of $D^0$ being a maximum of $V_{{\rm 1-loop}}^{(D)}$
  is not uncommon in supersymmetric field theories: 
 in fact, at the tree level, $V_{{\rm 1-loop}}^{(D)}$ has a maximum at $D^a =0$ in our case.
 The stability criterion is given for the extire effective potential
  with respect to the field space of the scalar vevs 
 and the vacuum we have found is a local minimum. 
Following a discussion on the estimation of the lifetime of the metastable vacuum in \cite{ISS}, 
one can show that
our metastable SUSY breaking vacuum can be made long-lived 
by taking the parameter $m/\Lambda^2$ to be sufficiently small:
 an estimate of its decay rate can be given, based on eqs. (\ref{vcond1}) and (\ref{vcond2}) in section 6.
 Space does not permit us to go into the detail here.



\section{finding an expansion parameter} 

The idea of the Hartree-Fock approximation is that the one-loop contributions become sufficiently large
 and start competing against the tree ones, eventually developing into a new vacuum.
 In fact, the gap equation has been obtained by matching
  these two. To make understanding on the validity of the approximation and proper applications to
   observables,  it is desirable to find an expansion parameter.
Let us regard the nonvanishing matrix elements of ${\cal F}, g$ and 
$W$ and their derivatives to be ${\cal O}(N^2)$. All three terms in the action can be rescaled to
have $N^2$ in front and $\frac{1}{N^2}$ may replace the original loop expansion parameter $\hbar$,
 becoming a new expansion parameter. Here, we will just demonstrate that both the first and the second terms
of the gap equation are ${\cal O}(N^2)$ and the present approximation is justified in the large $N^2$ limit.

For that purpose, we get back to recast the gap equation into another form with integrations over loop momenta. 
Recall that the part of the one-loop effective action (1PI vertex functional) which contains the auxiliary field
 $D^0$ is 
\ba
i \Gamma^{(D)}(\langle \phi_a \rangle, \langle \bar{\phi}_a \rangle, D^0) 
= \frac{i}{2} \int d^4x g_{00}(D^0)^2 + \frac{1}{2} \sum_a \int d^4x \int \frac{d^4k}{(2\pi)^4} \ln {\rm Det} {\cal D}_a \!\!\!\!\!\!/ \hspace*{2mm}(k, M_F, \bar{M}_F) 
\label{1PI}
\ea
where 
$
{\cal D}_a \!\!\!\!\!\!/ \hspace*{2mm}= \gamma^\mu k_\mu - {\cal M}_{F,a}, \quad
{\cal M}_{F,a} = 
\left(
\begin{array}{cc}
M_{F,a} & 0 \\
0 & \overline{M}_{F,a} \\
\end{array}
\right). 
$
After diagonalizating ${\cal M}_{F,a}$, we obtain
\ba
{\cal D}_{a,(diag)} \!\!\!\!\!\!\!\!\!\!\!\!\!\!\!\!\!\!\!/ \hspace*{11mm}= 
\left(
\begin{array}{cc}
\gamma \cdot k -|m_a|\lambda^{(+)} & 0 \\
0 & \gamma \cdot k -|m_a|\lambda^{(-)} \\
\end{array}
\right)
\ea
 $m_a=\langle g^{aa} \partial_a \partial_a W \rangle$. 
The gap equation reads 
\ba
0 = i \langle g_{00} \rangle -\frac{1}{2} \sum_a |m_a| \int \frac{d^4k}{(2\pi)^4} {\rm tr} 
\left[
\frac{1}{{\cal D}_{a,(diag)}\!\!\!\!\!\!\!\!\!\!\!\!\!\!\!\!\!\!/ } \hspace*{10mm} \frac{1}{D^0} \frac{\partial}{\partial D^0} 
\left(
\begin{array}{cc}
\lambda^{(+)} & 0 \\
0 & \lambda^{(-)} \\
\end{array}
\right)
\right]. 
\ea
 Here, the inclusion of the contribution from the counterterm at $\langle g_{00} \rangle$ is understood.
After some calculations, we obtain
\ba
 0 =  i \langle g_{00} \rangle -\frac{1}{2} \sum_a \langle g^{aa} {\cal F}_{aa0} \rangle  \langle \overline{{\cal F}_{0aa} g^{aa}} \rangle  
\int \frac{d^4k}{(2\pi)^4} 
\frac{k^2 - |m_a|^2 \frac{\Delta^2}{4}}{(k^2 - |m_a|^2 \lambda^{(+)2})(k^2 - |m_a|^2 \lambda^{(-)2})}. 
\label{gapeq}
\ea
In the unbroken phase of the $U(N)$ gauge group,
$\langle g_{aa,0} \rangle = \langle g_{00,0} \rangle$,
$\langle {\cal F}_{0aa} \rangle = \langle {\cal F}_{000} \rangle$, 
$\langle \partial_a \partial_a W \rangle = \langle \partial_0 \partial_0 W \rangle$, so that $m_a=m_0$. 
Eq. (\ref{gapeq}) is actually
\ba
0= i\langle g_{00} \rangle - \frac{N^2}{2} \langle g^{00} {\cal F}_{000} \rangle  \langle \overline{{\cal F}_{000} g^{00}} \rangle  
\int \frac{d^4k}{(2\pi)^4} 
\frac{k^2 - |m_0|^2 \frac{\Delta^2}{4}}{(k^2 - |m_0|^2 \lambda^{(+)2})(k^2 - |m_0|^2 \lambda^{(-)2})}. 
\label{gapequnbroken}
\ea
The second term is ${\cal O}(N^2)$, which is interpreted that an index loop $a= 0 \cdots N^2-1$
circulates together with the loop momenta.


\section{non-vanishing F term induced by $\langle D^0 \rangle \neq 0$ and 
its implications in fermion masses}

The above analysis demonstrates that dynamical supersymmetry beaking of
the ${\cal N} =1$ supersymmetry of the theory given
 by eq.(\ref{KtauW}) or 
 eq. (\ref{FIS'}) is triggered by the non-vanishing value of the $\langle D \rangle$ term
  to the leading order in the present approximation.
 For proper physical assessments and applications to observables, this is not sufficient however.
  As is already discussed in the introduction, 
  the nonvanishing VEV of  F term leads in principle to that of D term 
   and the converse can also be true.  let us see how this takes place in the present situation. 
    For illustration,  we will discuss its implications in the fermion mass spectrum.
 
In the vacuum of nonvanishing $\Delta$, 
the VEV's of the scalar fields get in fact shifted and another order parameter $\langle F^0 \rangle$ becomes nonvanishing 
as well as a result of the vacuum condition. 
The entire effective  potential up to one-loop to be extremized is
\ba
V = g^{ab} \partial_a W \overline{\partial_b W} -\frac{1}{2} g_{ab} D^a D^b + V_{{\rm 1-loop}} 
  + V_{{\rm c.t.}}. 
\ea
Let $\delta$ be a holomorphic variation of $V$ with respect to the scalar fields. 
In the unbroken phase of the vacua of $\Delta \ne 0$, 
\ba
\langle \delta V \rangle = -\langle \delta \phi^0 \rangle 
\left\{
\langle \partial_0 g_{00} \rangle |\langle F^0 \rangle|^2 + \langle \partial_0 \partial_0 W \rangle \langle F^0 \rangle 
+ \frac{1}{2} \langle \partial_0 g_{00} \rangle \langle D^0 \rangle^2 - \langle \partial_0 V_{{\rm 1-loop}} \rangle 
\right\}.
\ea
For the sake of simplicity, we assume here that the fourth prepotential derivatives vanish. Then
\ba
\langle \partial_0 V_{{\rm 1-loop}} \rangle = -2\langle g^{00} \rangle \langle \partial_0 g_{00} \rangle \langle V_{{\rm 1-loop}} \rangle.
\ea
The vacuum condition $\langle \delta V \rangle=0$ reads
\ba
|\langle F^0 \rangle|^2 + \frac{m_0}{\langle g^{00} \partial_0 g_{00} \rangle} \langle F^0 \rangle 
+ \frac{1}{2} \langle D^0 \rangle^2 
+ 2 \langle g^{00} \rangle \langle V_{{\rm 1-loop}} \rangle =0. 
\label{vcond1}
\ea
Combining with $\langle \overline{\delta} V \rangle=0$, we further obtain
\ba
\frac{m_0}{\langle g^{00} \partial_0g_{00} \rangle} \langle F^0 \rangle = \frac{m_0^*}{\langle \overline{g^{00} \partial_0 g_{00}} \rangle} \langle \bar{F}^0 \rangle. 
\label{vcond2}
\ea
Eqs. (\ref{vcond1}), (\ref{vcond2}) determine the value of non-vanishing F term triggered by the non-vanishing
D term.

 One implication of this phenomenon is that it is no longer true that the masses
 of the $SU(N)$ fermions to the leading order  are determined by
   the second derivative of the superpotential and the nonvanishing $\langle D^0 \rangle$:
   the determinations require the non-vanishing value of $\langle F^0 \rangle$ as well.
   In fact, the holomorphic part of the fermion mass matrix to the leading order is 
\ba
{\cal L}_{{\rm mass}}^{(holo)} &=& -\frac{1}{2} \langle g_{0a,a} \rangle \langle \bar{F}^0 \rangle \psi^a \psi^a 
+ \frac{i}{4} \langle {\cal F}_{0aa} \rangle \langle F^0 \rangle \lambda^a \lambda^a
-\frac{1}{2} \langle \partial_a \partial_a W \rangle \psi^a \psi^a 
+ \frac{\sqrt{2}}{4} 
\langle {\cal F}_{0aa} \rangle \psi^a \lambda^a 
\langle D^0 \rangle  \nonumber \\
&\equiv& -\frac{1}{2} \sum_{a=1}^{N^2-1} \Psi(x)^{a~t} M_{a,a} \Psi^a(x), \qquad 
\Psi^a(x) = 
\left(
\begin{array}{c}
\lambda^a(x) \\
\psi^a(x) \\
\end{array}
\right).  
\ea

   As for gaugino and matter fermions in the $U(1)$ sector, the hidden sector where the Nambu-Goldstone fermion
    resides,  the index loop circulates in the one-loop self energy part as well, which is,
    in additon to the above contributions, regarded as the leading contribution to the mass matrix. 
   The massless fermion ensured by the theorem
   is an admixture of $\lambda^{0}$ and $\psi^{0}$.

\section{application}

Finally, let us touch upon an application of our dynamical mechanism to
 supersymmetric particle physics phenomenology.  
The MSSM (minimal supersymmetric standard model) lagrangian symbolically reads 
${\cal L}_{MSSM} = {\cal L}_{{\rm gauge}} + {\cal L}_{{\rm matter}} + {\cal L}_{{\rm gauge-matter}}$.
Here the three terms represent respectively the part containing 
 the vector superfields of the gauge supermultiplets, 
 the one containing the chiral superfields of the matter supermultiplets 
 and the coupling of these two types of superfields. 
Clearly, the simplest and the most conservative application is to extend just the first term 
 to the type of actions discussed in this letter, 
 so that we do not introduce the mirror fermions of the standard model 
 which make theory non-chiral and which tend to endanger the asymptotic freedom of QCD, 
 giving rise to a Landau pole.
 As for the gauge group of this term, 
 we take, for instance, the product of a hidden gauge group 
 and the standard model (SM) gauge group $G^{\prime} \times G_{SM}$ and $G^{\prime}$ 
 includes the overall $U(1)$ responsible for the dynamical supersymmetry breaking. 
Due to the non-Lie algebraic nature of the third prepotential derivatives 
 ${\cal F}_{abc}$ or $\tau_{abc}$, we do not really need messenger superfields.

The supersymmetry breaking which originates from this sector is then transmitted to 
the rest of the theory by higher order loop corrections.
Once the gaugino masses are generated by our dynamical mechanism, 
        the sfermion masses $m^2_{{\rm sf}}$ are generated in the next loop order and take 
the following form:
\ba
m^2_{{\rm sf}}  \sim \frac{C_i(R)\alpha_i}{\pi} m(\Delta)_{i}^2 
\log \left( \frac{(m_a)_i^2}{m(\Delta)^2_i} \right).
\ea
Here $m(\Delta)_i~(i = SU(3), SU(2), U(1))$ are the gaugino masses of $G_{{\rm SM}}$
 (they correspond with ${\bf \Lambda}_a^{(-)}$ in eq. (\ref{eigenvalue})),
$C_i(R)$ is the quadratic Casimir of representation $R$. 
This is a general feature common to gaugino mediation, 
 in particular, the proposal of \cite{FNW} and applies here as well. 
These scalar masses are positive and flavor-blind 
 and are free from the supersymmetric flavor and CP problems. 
 Furthermore, the generation mechanism of these masses themselves is insensitive to the UV scale
 \cite{FNW} and the window to the multi-TEV scale is confined to the sector 
 whose dynamics is discussed in this letter.
 
The authors' research  is supported in part by the Grant-in-Aid for Scientific Research  
  from the Ministry of Education, Science and Culture, Japan (23540316 (H.~I.), 21244036 (N.~M.))
  and by Keio Gijuku Academic Development Funds (N.~M.).

 
\end{document}